\documentclass[12pt]{iopart}

\usepackage{graphicx} 
\usepackage{dcolumn}  
\usepackage{bm}       

\begin{document}

\title{Algebraic solution of a graphene layer in a transverse
electric and perpendicular magnetic fields}

\author{N. M. R. Peres$^1$ and Eduardo V. Castro$^2$}

\address{$^1$Center of Physics and Department of
  Physics, University of Minho, P-4710-057, Braga, Portugal}

\address{$^2$ CFP and Departamento de F\'{\i}sica, 
  Faculdade de Ci\^encias Universidade do Porto, P-4169-007 Porto, Portugal}

\begin{abstract}
We present an exact algebraic solution of a single graphene plane
in transverse electric and perpendicular magnetic fields.
The method presented gives both the eigen-values and the
eigen-functions of the graphene plane.
It is shown that the eigen-states of the problem can be casted
in terms of  coherent states, which  appears in a natural way
from the formalism.
\end{abstract}

\submitto{\JPCM}

\section{Introduction}

A major breakthrough in condensed matter physics took place 
when K. S. Novoselov {\it et al.} \cite{NGM+04} , at Manchester
University, UK,
discovered
an electric field effect in atomically thin carbon films.
This electric field effect is characterized by the control of
the electronic density in the films using a backgate
setup. These atomically thin carbon films were thought not to
exist since long range order in two dimension can not occur.
The system solves the apparent paradox by forming ripples.
A single atomic layer of these thin carbon films is called graphene
and its electric and magneto-electric properties triggered
a new research field in condensed matter physics. 
The manufacture of graphene was followed by the production
of other 2D crystals\cite{pnas}, which, however, have not the same
exciting properties as graphene does. Applying high magnetic fields
to a graphene sample,
the Manchester group discovered that the quantization rule
for the Hall conductivity is not the same one observes in the 
two-dimensional electron gas, being given instead by\cite{NGM+05}

\begin{equation}
\sigma_{\rm Hall} = 4\left(n+\frac 1 2\right)\frac {e^2}h\,,
\label{sigmahall}
\end{equation}

with $n$ an integer including zero. 
A confirmation of this result was independently
obtained by Philip Kim's group\cite{ZTS+05}, at Colombia University,
New York, USA. This
new quantum Hall effect was predicted by two groups working independently
and using different methods \cite{PGN06,GS05}. As explained by the two groups
the new quantization rule for the Hall conductivity is a consequence
of the dispersion relation of the electrons in the honeycomb lattice. 
This dispersion resembles the spectrum of ultra relativistic particles, 
i.e., the dispersion is that of 
particles of zero rest mass and an effective velocity of light. For graphene
the effective velocity of light is  $v_F=c/300$, with $c$ the true
velocity of light.

For a qualitative
description of the physics of graphene, both
theoretical and experimental, see references \cite{castro}
by Castro Neto  {\it et al.}, \cite{katsnelson} by M. I.~Katsnelson,
and \cite{Nov07} by  Geim and Novoselov.

\begin{figure}
\begin{center}\includegraphics*[width=8cm]{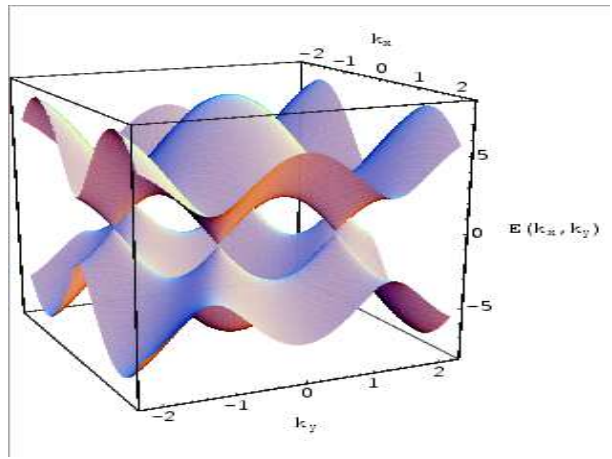}
\end{center}
\caption{\label{Fig_dispersion} Valence and conduction bands
of electrons in graphene. The two bands touch each other
in six points of the Brillouin zone, called Dirac points.}
\end{figure}

\begin{figure}
\begin{center}\includegraphics*[width=8cm]{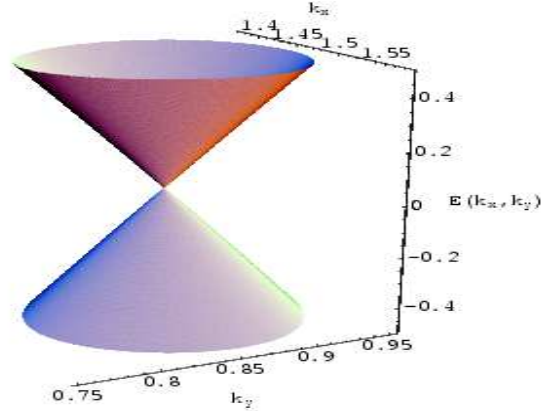}
\end{center}
\caption{\label{Fig_cone} Valence and conduction bands
close to one of the six Dirac points. It is clear that
the bands can be approximated by a conical dispersion.}
\end{figure}

In Figure \ref{Fig_dispersion}  we show the energy dispersion
of electrons in the honeycomb lattice. The spectrum shows 
a valence (lower) and a conduction (upper) bands.
Since graphene has one electron
per unit cell the valence band is completely filled and the properties
of the system are determined by the nature of its spectrum close to the
points where the valence and the conduction bands touch each other.
These points are called Dirac points and are in number of six.
In Figure \ref{Fig_cone} we show the spectrum close to the Dirac
points. It is clear that the spectrum has conical shape of the form
\begin{equation}
E=\pm v_Fp, 
\end{equation}
where $p$ is the magnitude of the momentum $\bm p$
around the Dirac point. 

A relativistic particle has an energy given
by
\begin{equation}
E=\sqrt{m^2c^4+p^2c^2}\,,
\end{equation}
and therefore an ultra-relativistic particle ($m\rightarrow 0$) has a spectrum
given by
\begin{equation}
E=cp\,.
\label{ur}
\end{equation}
It is clear from equation (\ref{ur}) that  electrons
in graphene, close to the Dirac points, 
have an energy dispersion with a formal equivalence
to ultra-relativistic particles.
As a consequence the quantum properties of  
the system has to be described by the massless (ultra-relativistic)
Dirac equation in two plus one dimensions.

We are interested in
studying the spectrum of massless Dirac particles
in the presence of a magnetic field perpendicular
to the plane and an in plane homogeneous electric field, both static,
a situation that occurs in the Hall effect. In the next section
we present a full algebraic solution to this quantum problem.

\section{Algebraic solution}

\subsection{Hamiltonian}
The massless Dirac equation in two plus one dimensions
has the form
\begin{equation}
v_F (\sigma_x p_x + \sigma_y p_y)\Psi(\bm r,t)=
i\hbar \frac {\partial \Psi(\bm r,t)}{\partial t}\,,
\end{equation}
where $\sigma_i$, $i=x,y,z$, represents the Pauli spin matrices and
$p_i$, $i=x,y$, is the $i$ component of the momentum operator 
in the position basis
 $\bm p =-i\hbar \bm \nabla$. Since we are looking for stationary states
we make the substitution $\Psi(\bm r,t)=\psi(\bm r)e^{-i\epsilon t/\hbar}$. 
This substitution transforms the Dirac equation into an eigenvalue
problem of the form
\begin{equation}
v_F (\sigma_x p_x + \sigma_y p_y)\psi(\bm r)=\epsilon\psi(\bm r)\,.
\label{HpEp}
\end{equation}
The introduction of a magnetic field into a quantum mechanical
problem is made by transforming the momentum operator according
to the rule (minimal coupling) 
\begin{equation}
\bm p \rightarrow \bm p - q\bm A(\bm r), 
\end{equation}
where $\bm A(\bm r)$
is the vector potential and $q$ is the charge of the particle. 
The magnetic field $\bm B$ is obtained
from $\bm A$ using the relation $\bm B =\bm \nabla\times \bm A$.
There is a lot of freedom in choosing $\bm A$ and a common choice,
known as Landau gauge, is $\bm A=(-By,0,0)$. Let us now assume that 
in addition to the magnetic field one has a homogeneous
electric field, perpendicular
to the magnetic field, and oriented along the $y$ direction. This
adds to the Hamiltonian a term of the form
\begin{equation}
qV\bm 1 = q\mathcal{E}y\bm 1\,,
\end{equation}
where $V$ is the electric potential associated with the applied
electric field ${\bm E} = (0,\mathcal{E},0) $ and
$\bm 1$ is the $2\times 2$ unit matrix.

Putting all together, the problem of a graphene layer 
in the presence of a  magnetic
field perpendicular to the  layer and of an electric field
parallel to the layer has its Hamiltonian, in the position basis,
given by:
\begin{equation}
H(x,y) = v_F 
\left(\begin{array}{cc}
q\mathcal{E}y/v_F & p_x-ip_y+qBy\\
p_x+ip_y+qBy & q\mathcal{E}y/v_F
\end{array}\right)
\,.
\label{H1}
\end{equation}

The eigenproblem $H(\bm r)\psi(\bm r)=\epsilon\psi(\bm r)$ 
can be further simplified by
representing the eigenfunction $\psi(\bm r)$ as
\begin{equation}
\psi(\bm r) = e^{ikx}\phi(y)\,,
\label{psi}
\end{equation}
suggested by the translational invariance of Eq.~(\ref{H1})
along the $x$ direction.
Because we are dealing with electrons one has $q=-e$, with $e>0$.
Using equation (\ref{psi}) in equation (\ref{H1}) we obtain
\begin{equation}
 v_F 
\left(\begin{array}{cc}
-e\mathcal{E}y/v_F  & k\hbar-ip_y-eBy\\
k\hbar+ip_y-eBy & -e\mathcal{E}y/v_F 
\end{array}\right)
\phi(y)=\epsilon\phi(y)\,.
\label{H2}
\end{equation}
Next we perform a change of variables $ y = \bar y l_B+l^2_Bk$
and $\partial/\partial \bar y = l_B \partial/\partial y$
(corresponding to the introduction of the adimensional length $\bar y$),
with $l_B=\sqrt{\hbar/(eB)}$ the magnetic length,
and introduce the operators
\begin{eqnarray}
a&=& \frac {1}{\sqrt 2}(\hat{\bar{y}} + \partial/\partial \hat{\bar y})\,,
\label{eq:ladder1}\\
a^\dag&=&\frac {1}{\sqrt 2}(\hat{\bar{y}} - \partial/\partial \hat{\bar y})\,,
\label{eq:ladder2}
\end{eqnarray}
which satisfy the standard commutation relation $[a,a^\dag]=1$.
Note that in Eqs.~(\ref{eq:ladder1}) and (\ref{eq:ladder2}) we have 
used the hat to distinguish between operators and their matrix elements 
in a given basis.

Performing standard manipulations the Hamiltonian operator can be brought
into a more transparent form
\begin{equation}
\hat{H} =- 
\left(\begin{array}{cc}
  e\mathcal{E}l_B^2k  + E_B(a+a^\dag)& E_Fa \\
E_Fa^\dag &  e\mathcal{E}l_B^2k+ E_B(a+a^\dag)
\end{array}\right)
\,,
\end{equation} 
with $E_F=\sqrt 2 v_F\hbar /l_B$ and $E_B=e\mathcal{E}l_B/\sqrt 2$.
The eigenvalue equation one needs to solve has the form
\begin{equation}
\left(\begin{array}{cc}
 E_B(a+a^\dag)& E_Fa \\
E_Fa^\dag &  E_B(a+a^\dag)
\end{array}\right)
\left(\begin{array}{c}
\vert a_1\rangle\\
\vert a_2\rangle
\end{array}\right)
= \epsilon_0
\left(\begin{array}{c}
\vert a_1\rangle\\
\vert a_2\rangle
\end{array}\right)
\,,
\label{eq_H3}
\end{equation}
where $\epsilon_0=-(\epsilon +e\mathcal{E}l_B^2k)$.
The eigenproblem is now in its most simplified form, with an effective
Hamiltonian operator given by,
\begin{equation}
\hat{\mathcal{H}} = 
\left(\begin{array}{cc}
 E_B(a+a^\dag)& E_Fa \\
E_Fa^\dag &  E_B(a+a^\dag)
\end{array}\right)
\label{eq:Heff}
\end{equation}

\subsection{Diagonalization method}
\label{sec:diagmethod}

Before going further with the diagonalization
it is worth mentioning some 
properties of the Hamiltonian operator $\hat{\mathcal H}$, defined in 
Eq.~(\ref{eq:Heff}). First we define $\hat{\bar{\mathcal H}}$ as,
\begin{equation}
\hat{\bar{\mathcal{H}}}=\sigma_{z}\hat{\mathcal{H}}\sigma_{z}=
\left(\begin{array}{cc}
E_{B}(a+a^{\dagger}) & -E_{F}a\\
-E_{F}a^{\dagger} & E_{B}(a+a^{\dagger})
\end{array}\right)
,
\label{eq:Hbar}
\end{equation}
which, by definition, is an operator acting on the same Hilbert space
as $\hat{\mathcal{H}}$.
Then we make the observation that both $\hat{\mathcal H} + 
\hat{\bar{\mathcal H}}$ and $\hat{\bar{\mathcal H}} \hat{\mathcal H}$ 
can be written as the $2 \times 2$ unit matrix times a simple operator 
(not a matrix operator) plus a $2 \times 2$ real matrix. 
The same holds for their linear combination, which enables us to write
\begin{equation}
\mu (\hat{\mathcal H} + \hat{\bar{\mathcal H}}) +
\nu \hat{\bar{\mathcal H}} \hat{\mathcal H}
= \hat J \bm 1 + \bm K,
\label{eq:manipHHbar}
\end{equation}
for some simple operator $\hat J$, some $2 \times 2$ real matrix $\bm K$,
and real $\mu$ and $\nu$.
Now let the spinor $\vert \psi\rangle$ be an eigenstate of $\hat{\mathcal H}$. 
Applying the left hand member of Eq.~(\ref{eq:manipHHbar}) to 
$\vert \psi\rangle$ we obtain,
\begin{equation}
[\mu(\hat{\mathcal{H}}+\hat{\bar{\mathcal{H}}})+
\nu\hat{\bar{\mathcal{H}}}\hat{\mathcal{H}}]
\vert \psi\rangle
=(\mu \epsilon_{0}+\mu\hat{\bar{\mathcal{H}}}+
\nu \epsilon_{0}\hat{\bar{\mathcal{H}}})
\vert \psi\rangle,
\label{eq:munu}
\end{equation}
which means that if we chose $\mu=\epsilon_{0}$ and $\nu=-1$ we
reduce our problem to,
\begin{equation}
(\hat{J}\bm{1}+\bm{K}) \vert \psi\rangle
= \epsilon_{0}^{2} \vert \psi\rangle,
\label{eq:simple}
\end{equation}
with
\begin{equation}
\hat{J}=(E_{F}^{2}-2E_{B}^{2})\hat{n}-E_{B}^{2}(aa+a^{\dagger}a^{\dagger}) +
2\epsilon_{0}E_{B}(a+a^{\dagger}),
\label{eq:J}
\end{equation}
where $\hat n = a^{\dagger}a$ is the number operator, and
\begin{equation}
\bm{K}=
\left(\begin{array}{cc}
E_{F}^{2}-E_{B}^{2} & E_{F}E_{B}\\
-E_{F}E_{B} & -E_{B}^{2}
\end{array}\right)
.
\label{eq:K}
\end{equation}

Equation~(\ref{eq:simple}) is indeed simpler than our starting point, 
Eq.~(\ref{eq_H3}). We readily see that the spinor $\vert \psi\rangle$,
given by
\begin{equation}
\vert\psi\rangle=
\left(\begin{array}{c}
\vert a_1\rangle\\
\vert a_2\rangle
\end{array}\right)
\,,
\end{equation}
can be written as
\begin{equation}
\vert\psi\rangle=\vert \phi\rangle
\left(\begin{array}{c}
u\\
v
\end{array}\right)
\,,
\end{equation}
where $\vert \phi\rangle$ is the eigenvector of
the operator $\hat J$ and the spinor 
\begin{equation}
\chi^R=
\left(\begin{array}{c}
u\\
v
\end{array}\right)
\,,
\end{equation}
is the right eigenvector of the eigenvalue problem
\begin{equation}
\bm K \chi^R = \lambda \chi^R
\,.
\label{eq_eigen_lambda}
\end{equation}
Nevertheless, there is one subtlety we must consider.
Our simpler eigenproblem defined by Eq.~(\ref{eq:simple}) is such that
their eigenvalues are $\epsilon_0^2$, the squared eigenvalues of the original
problem given by Eq.~(\ref{eq_H3}). If $\epsilon_0$ in 
Eq.~(\ref{eq_H3}) and $\epsilon_0^2$ in Eq.~(\ref{eq:simple}) have the
same degeneracy it is guaranteed that eigenvectors of both problems
are the same, and we can carry on the diagonalization with any
of those equations. If, however, both $+ \epsilon_0$ and $- \epsilon_0$
are eigenvalues of $\hat{\mathcal{H}}$ in Eq.~(\ref{eq_H3}), then our
simple eigenproblem in Eq.~(\ref{eq:simple}) will show an extra double
degeneracy. This extra degeneracy must be handled carefully.
Due to mixing the
corresponding degenerate eigenvectors need not to be eigenstates of
$\hat{\mathcal{H}}$, and these  have to be found as
particular linear combinations of the degenerated eigenvectors. 
It is instructive to
switch off the electric field, for which Eq.~(\ref{eq_H3}) can be easily
solved\cite{PGN06}, and check whether the above problem shows up in the present
case.

\subsection{Zero electric field case}
\label{sec:noE}

In the absence of the electric field we have $\mathcal{E}=E_B=0$. 
As a consequence the 
operator $\hat J$ in Eq.~(\ref{eq:J}) is already in its diagonal form, 
being analogous to the 1D harmonic oscillator:
 $\hat{J}\vert n\rangle = E_F^2 n \vert n\rangle$.
In what regards the eigenproblem defined by Eq.~(\ref{eq_eigen_lambda}) 
for $\mathcal{E} = 0$, we can obtain the corresponding eigenvectors as
\begin{equation}
\chi^R_+ =
\left(\begin{array}{c} 
1\\0
\end{array}\right)
\hspace{0.5 cm}
\textrm{and}
\hspace{0.5 cm}
\chi^R_- =
\left(\begin{array}{c} 
0\\1
\end{array}\right)
\,,
\label{eq:chiNoE}
\end{equation}
with eigenvalues
$\lambda_+ = E_F^2$ and $\lambda_- = 0$, respectively.
Recalling Eq.~(\ref{eq:simple}) we get
$\epsilon_0^2(n,\pm) = E_F^2n+\lambda_{\pm}$,
where we recognize immediately the double degeneracy
$\epsilon_0^2(n,+)=\epsilon_0^2(n+1,-)=E_F^2(n+1)$.
This degeneracy is in fact due to the presence of both $+\epsilon_0$ and
$-\epsilon_0$ as eigenvalues of $\hat{\mathcal{H}}$ in the absence of electric
field. Solving Eq.~(\ref{eq_H3}) directly for $\mathcal{E} = 0$ 
gives\cite{PGN06}
$\epsilon_0 = \pm E_F\sqrt{n+1}$ in addition to the zero eigenvalue 
$\epsilon_0 = 0$, with eigenvectors
\begin{equation}
\vert \psi\rangle_\pm = 
\left(\begin{array}{c}
\vert n\rangle\\ \pm \vert n + 1\rangle
\end{array}\right)
\hspace{0.5 cm}
\textrm{and}
\hspace{0.5 cm}
\vert \psi\rangle_0 =
 \left(\begin{array}{c}
0\\ \vert 0\rangle
\end{array}\right)
\,,
\label{eq:LsNoE}
\end{equation}
respectively for nonzero and zero eigenvalues.
Therefore we see that our method gives correctly $\vert \psi\rangle_0$,
the only non-degenerate eigenvector,
\begin{equation}
\vert \psi\rangle_0 = \vert 0 \rangle \chi^R_-\,,
\label{eq:psichiNoE1}
\end{equation}
while $\vert \psi\rangle_\pm$ is given as the bonding and anti-bonding
combination of degenerate eigenvectors $\vert n\rangle \chi^R_+$
and $\vert n + 1\rangle \chi^R_-$:
\begin{equation}
\vert \psi\rangle_\pm = \vert n\rangle \chi^R_+ \pm \vert n + 1\rangle \chi^R_-
\,.
\label{eq:psichiNoE2}
\end{equation}
As extra degeneracies due to the presence of a finite electric field
are not to be expected, we will be able to identify any double degeneracy
arising from Eq.~(\ref{eq:simple}) as a consequence of the presence of 
symmetrical eigenvalues $\pm \epsilon_0$ in the original problem.

\subsection{Finite electric and magnetic fields}
With the above analysis 
in mind we proceed with the diagonalization of our problem
in a finite electric and magnetic fields using Eq.~(\ref{eq:simple}). 
Let us start by solving 
Eq.~(\ref{eq_eigen_lambda}). The corresponding eigenvalues are given by,
\begin{equation}
\lambda_\pm = -E_B^2 +
\frac{1}{2}\left(E_{F}^{2} \pm E_{F}\sqrt{E_{F}^{2}-4E_{B}^{2}}\right)\,,
\label{eq:lambda}
\end{equation}
and as right eigenvectors we obtain,
\begin{equation}
\chi^R_\pm =
\sqrt{\left|\frac{E_{B}}{E_{F}}\right|}
\left(\begin{array}{c}
-\sqrt{C_\pm}\\
1/\sqrt{C_\pm}
\end{array}\right)
\,,
\label{eq:chiR}
\end{equation}
with 
\begin{equation}
C_\pm = \frac 1 2 \left(E_F/|E_B| \pm \sqrt{E_{F}^{2}/E_{B}^{2}-4}\right) \,.
\label{eq:Cpm}
\end{equation}
From equation (\ref{eq:lambda}) we see that 
$E_F$ and $E_B$ must satisfy the relation,
\begin{equation}
E^2_F \ge 4E^2_B \,,
\label{eq:breakdown}
\end{equation}
if real eigenvalues are to be obtained. The meaning of this inequality
is discussed later.
Having solved the eigenproblem (\ref{eq_eigen_lambda}), 
 the eigenproblem
\begin{equation}
\hat J \vert \phi \rangle  = 
(\epsilon_0^2 -\lambda_\pm) \vert \phi \rangle
\label{eq_eigen_a}
\end{equation}
remains to be solved.
The solution of the eigenproblem (\ref{eq_eigen_a}) is obtained
in  three steps. 
First we write the operator $\hat J$
as the sum of two terms, $ \hat{H}_1 + \hat{H}_2$, given by
\begin{equation}
\hat{H}_1 = (E^2_F-2E^2_B)\hat n - E^2_B(aa+a^\dag a^\dag)\,,
\label{H11}
\end{equation}
and
\begin{equation}
\hat{H}_2 = 2\epsilon_0E_B(a+a^\dag)\,.
\label{H22}
\end{equation}
As a second step we diagonalize the Hamiltonian $\hat{H}_1$ using the 
canonical transformation
\begin{equation}
a^\dag = \cosh U \gamma^\dag - \sinh U \gamma\,,
\end{equation}
and the corresponding Hermitian conjugated form for $a$. Replacing 
the $a^\dag$ and the $a$ operators in  (\ref{H11}) one obtains
\begin{eqnarray}
\hat{H}_1 &=& E^2_B2\sinh U \cosh U + (E_F^2-2E^2_B)\sinh^2U\nonumber
     +(\gamma^\dag\gamma^\dag + \gamma\gamma)
[-E^2_B(\cosh^2U+\sinh^2U)\nonumber\\
&-& (E^2_F-2E^2_B)\sinh U\cosh U]
+\gamma^\dag\gamma [(E_F^2-2E^2_B)(\cosh^2 U+\sinh^2U)\nonumber\\
&+&4E_B^2\sinh U\cosh U]\,. 
\end{eqnarray}
In order for $\hat{H}_1$ to be diagonal we require that the coefficient
multiplying the $(\gamma^\dag\gamma^\dag + \gamma\gamma)$ term
should be null, leading to 
\begin{equation}
[-E^2_B(\cosh^2U+\sinh^2U)
- (E^2_F-2E^2_B)\sinh U\cosh U]=0
\end{equation}
which can be cast in the form
\begin{equation}
\tanh (2U) =-\frac {2E^2_B}{E^2_F-2E^2_B}\,.
\label{U}
\end{equation}
We note that since $\cosh U>0$ for any value of $U$ one must
have $\sinh U<0$ in order to satisfy equation (\ref{U}).
The result (\ref{U}) together with $\cosh^2U-\sinh^2U=1$
can be recasted in the form 
\begin{equation}
\sinh^2 U = -\frac 1 2 \left[1-
(E^2_F-2E^2_B)/\omega
\right]\,,
\end{equation}
and 
\begin{equation}
\cosh^2 U = \frac 1 2 \left[1+
(E^2_F-2E^2_B)/\omega
\right]\,,
\end{equation}
leading to 
\begin{equation}
\sinh U\cosh U = -\frac {E^2_B}{\omega} 
\end{equation}
with
$\omega=\sqrt{E^4_F-4E^2_FE^2_B}$
and $E^2_F>4E^2_B$. 
Using the results for $\sinh U$ and $\cosh U$
one can  write the piece $\hat{H}_1$ of the full Hamiltonian as
\begin{equation}
\hat{H}_1 = \frac 1 2 [ \omega - (E^2_F-2E^2_B)]+
 \omega\gamma^\dag\gamma \equiv C_1 + \omega\gamma^\dag\gamma\,,
\end{equation} 
with $C_1 = [ \omega - (E^2_F-2E^2_B)]/2$.
The piece $\hat{H}_2$ has now the form 
\begin{equation}
\hat{H}_2 =2\epsilon_0E_B (\cosh U-\sinh U)(\gamma^\dag+\gamma)\equiv
C_2(\gamma^\dag+\gamma)\,,
\end{equation}
where $C_2 = 2\epsilon_0E_B (\cosh U-\sinh U)$.
The third step requires the diagonalization of $\hat{H}_1+\hat{H}_2$ in the 
new form, written in terms of the $\gamma-$operators;
this is accomplished by the transformation $\gamma^\dag = \beta^\dag + Z$
(with $Z=-C_2/\omega$),
leading to
\begin{equation}
\hat{H}_1+\hat{H}_2=C_1-\frac {C_2^2}{\omega} + \omega \beta^\dag \beta\,,
\label{eq_h1_h2}
\end{equation}
which has the desired diagonalized form. The eigenenergies of
the Hamiltonian (\ref{eq_h1_h2}) have the form, 
\begin{equation}
\omega_n = C_1-\frac {C_2^2}{\omega} + \omega  n
= \frac{1}{2}[\omega-(E_{F}^{2}-2E_{B}^{2})]-\frac{4\epsilon_{0}^{2}E_{B}^{2}E_{F}^{2}}{\omega^{2}}+\omega n\,,
\end{equation}
and the ground state obeys the relation 
\begin{equation}
\beta \vert 0;\beta\rangle=0 \Leftrightarrow \gamma\vert 0;\beta\rangle = 
Z   \vert 0;\beta\rangle.
\label{eq_eigen_vector_beta}
\end{equation}
One should note that the state  $\vert 0;\beta\rangle$
is an eigenstate the $\gamma$ operator with eigenvalue $Z$; it is therefore
said that $\vert 0;\beta\rangle$ is a coherent state
of the operator $\gamma$. This last result allows us to write the vacuum
of the $\beta$ operators in terms of the vacuum of the $\gamma$
operators as 
\begin{equation}
 \vert 0;\beta\rangle = e^{Z\gamma^\dag}\vert 0;\gamma\rangle\,,
\end{equation} 
and any eigenstate is written in terms of the 
$\beta$-operators as
\begin{equation}
\vert n;\beta\rangle = \frac 1 {\sqrt {n !}}(\beta^\dag)^n \vert 0;\beta\rangle
= \frac 1 {\sqrt {n !}}\left(\gamma^\dag-Z\right)^n 
e^{Z\gamma^\dag}\vert 0;\gamma\rangle\,.
\end{equation}
The eigenenergies $\epsilon_0^2$ of our simpler eigenproblem defined
in Eq.~(\ref{eq:simple}) are obtained from [see Eq.~(\ref{eq_eigen_a})]
$\epsilon_0^2 - \lambda_\pm = \omega_n$, leading to
\begin{equation}
\epsilon_0^2(n,\pm)=\frac {\omega^3}{E_F^4}[n+(1 \pm 1)/2]\,.
\label{eq:e2}
\end{equation}
The double degeneracy 
$\epsilon_0^2 (n, +) = \epsilon_0^2 (n + 1, -)$ for $n \neq 0$
is immediately recognized. From the analysis we have made in 
Secs.~\ref{sec:diagmethod} and~\ref{sec:noE} it is now obvious that this 
degeneracy signals the presence of both solutions $\pm \epsilon_0$ in the
original problem [Eq.~(\ref{eq_H3})].
Moreover, as the eigenvectors of the finite electric field problem have
to equal Eq.~(\ref{eq:LsNoE}) when $\mathcal{E} \rightarrow 0$
we arrive at the following solution,
\begin{equation}
\epsilon(n,\pm)= -e\mathcal{E}l_B^2k \mp 
\frac {(E_F^2-4E_B^2)^{3/4}}{E_F^{1/2}}\sqrt{n+1}
\label{eq:LL}
\end{equation}
with eigenvectors given by
\begin{equation}
\vert \psi \rangle_\pm =
\sqrt{\left|\frac{E_{B}}{E_{F}}\right|}
\left(\begin{array}{c}
-\vert n; \beta \rangle \sqrt{C_+} \mp \vert n + 1; \beta \rangle \sqrt{C_-}\\
\vert n; \beta \rangle/\sqrt{C_+} \pm \vert n + 1; \beta \rangle /\sqrt{C_-}
\end{array}\right)
\,,
\label{eq:Ls}
\end{equation}
where $C_\pm$ is defined in Eq.~(\ref{eq:Cpm}). In addition there is a
single non-degenerate solution $\epsilon_0^2(0,-) = 0$,
which gives $\epsilon = -e\mathcal{E}l_B^2k$, and has as eigenvector
\begin{equation}
\vert \psi \rangle_0 =
\sqrt{\left|\frac{E_{B}}{E_{F}}\right|}
\left(\begin{array}{c}
-\sqrt{C_-}\\
1/\sqrt{C_-}
\end{array}\right)
\vert 0; \beta \rangle 
\,.
\label{eq:Lszero}
\end{equation}
This concludes our solution. The eigen-values (\ref{eq:LL})  
agree with those obtained
by Lukose {\it et al.}\cite{lukose}. These authors solved the
problem directly in the position basis by transforming the original
problem, by means of a Lorentz boost transformation, 
into a case where the electric field is null.

\subsection{Physical interpretation}
The standard 2D electron gas pierced by a perpendicular
magnetic field is known after Landau\cite{landau} to have a spectrum
given by,
\begin{equation}
\epsilon(n) = \hbar  \omega_\textrm{c} 
\Big(n + \frac 1 2 \Big),\hspace{0.5 cm} n=0,1,2,\dots
\label{eq:landau}
\end{equation}
in complete analogy with the quantum harmonic oscillator,
where $\omega_\textrm{c} =|eB|/m$ is the cyclotron frequency for electrons
with mass $m$. The so-called Landau levels are equally spaced with
level separation $\hbar \omega_c$, which increases linearly with $B$.
An in-plane electric field is easily handled by the transformation
$a^\dagger = b^\dagger + e\mathcal{E}l_B/(\hbar \omega_c \sqrt{2})$, whose
major consequence is a shift of the entire spectrum,
\begin{equation}
\epsilon(n) = - e\mathcal{E}l_B^2k - 
\frac{e^2\mathcal{E}^2l_B^2}{2\hbar \omega_c} + 
\hbar \omega_\textrm{c}\Big(n + \frac 1 2 \Big),\hspace{0.5 cm} n=0,1,2,\dots
\label{eq:landauE}
\end{equation}
with no change for the cyclotron frequency.

Landau levels in graphene are completely different from Landau levels
in standard 2D electron gas. As mentioned in Sec.~\ref{sec:noE} graphene's
spectrum in perpendicular magnetic field is given by\cite{PGN06},
\begin{equation}
\epsilon(n) = \pm \hbar \tilde{\omega}_\textrm{c} \sqrt{n},
\hspace{0.5 cm} n=0,1,2,\dots
\label{eq:Glandau}
\end{equation}
where the cyclotron frequency is 
$\tilde{\omega}_\textrm{c} = \sqrt{2}v_F / l_B$.
Two major differences become apparent when comparing Eqs.~(\ref{eq:Glandau})
 and~(\ref{eq:landau}). Firstly, Landau level spacing in graphene is not
constant due to the square root in Eq.~(\ref{eq:Glandau}). Secondly,
the standard 2D electron gas has $\omega_\textrm{c} \propto B$ whereas
graphene shows $\tilde{\omega}_\textrm{c} \propto \sqrt B$.
These dissimilarities are a direct consequence of
the effective ultra relativistic nature of the quasi-particles in
graphene.

\begin{figure}
\begin{center}\includegraphics*[%
  scale=0.5]{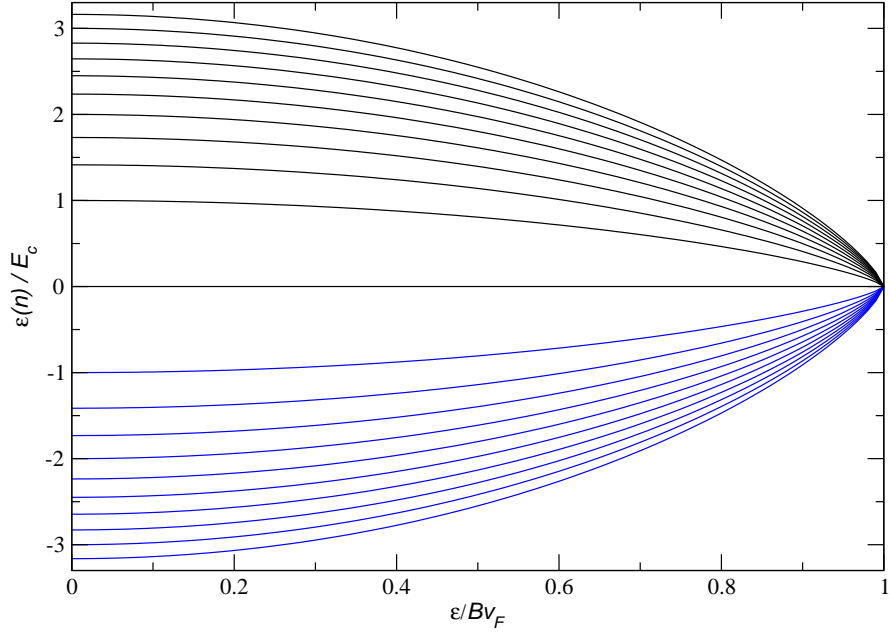}
\end{center}
\caption{\label{Fig_LL} First ten Landau levels as function of
$|\mathcal{E}|/ v_F |B|$. The quantity $E_c$ is the cyclotron energy
$\hbar\tilde\omega_c$. The momentum $k$ was chosen to be zero.}
\end{figure}

As regards the presence of an in-plane electric field and perpendicular
magnetic field in graphene we have shown that Landau levels are given by
Eq.~(\ref{eq:LL}), which can be cast in the form,
\begin{equation}
\epsilon(n)= -e\mathcal{E}l_B^2k \mp \hbar \Omega_\textrm{c} \sqrt{n},
\hspace{0.5 cm} n=0,1,2,\dots
\label{eq:GlandauE}
\end{equation}
where the new cyclotron frequency reads
\begin{equation}
\Omega_\textrm{c} = 
\sqrt{2} \frac{v_F}{l_B}  [1-\mathcal{E}^2/(B^2v_F^2)]^{3/4}\,.
\label{eq:cyclE}
\end{equation}
Thus, unlike the usual 2D electron gas, graphene's cyclotron
frequency is renormalized by the electric field, as can be seen
in Eq.~(\ref{eq:cyclE}), which, of course, reduces to $\tilde{\omega}_c$
in the limit $\mathcal{E}\rightarrow 0$.
More important though is the fact that $|\mathcal{E}| \leq v_F |B|$
must be realized if $\Omega_\textrm{c}$ is to be real. This last inequality
is exactly the same expressed in Eq.~(\ref{eq:breakdown}), and
its meaning is now unveiled. As $\mathcal{E}$ approaches $v_F B$ from
below $\Omega_\textrm{c}$ becomes smaller and smaller, and Landau levels
become closer and closer. Eventually, the electric field is such that
$|\mathcal{E}| = v_F |B|$, which implies $\Omega_\textrm{c} = 0$, and
consequent collapse of Landau levels. 
In Fig. \ref{Fig_LL} we show the first ten Landau levels (for positive
and negative energies) as function of $|\mathcal{E}|/ v_F |B|$;
the collapse of the Landau levels is clear.
For $|\mathcal{E}| > v_F |B|$
the present solution is  not valid.

\section{Concluding remarks}
The problem of a single graphene plane
in transverse electric and perpendicular magnetic fields 
assembles in a simple  way several
algebraic methods of diagonalizing bilinear problems.
The matrix form of the Dirac Hamiltonian~--~the low energy
effective Hamiltonian for graphene~--~calls for several
non-standard manipulations where canonical transformations
and the concept of coherent state appear in a natural way.
Furthermore, alike the standard 2D electron gas pierced by a magnetic field,
an additional in-plane electric field in graphene induces cyclotron
frequency renormalization. Moreover, when the electric field equals
the critical value $v_F B$ Landau level collapse is observed.

\subsection*{Acknowledgments}

E.V.C. was supported by FCT
through Grant  No.~SFRH/BD/13182/2003
and EU through POCTI (QCAIII).
N.M.R.P. is thankful to the ESF Science Programme No.~INSTANS~2005-2010
and FCT and EU under the Grant PTDC/FIS/64404/2006.


\end{document}